\documentclass[aps,twocolumn,longbibliography,showpacs,nofootinbib,numbers,openany]{revtex4-2}
\usepackage{graphicx}

\usepackage{inconsolata}
\usepackage{amssymb}
\usepackage{amsmath} 

\usepackage{multirow, colortbl, booktabs}
\usepackage[table,dvipsnames]{xcolor}

\definecolor{c_charged}{rgb}{1,0.85,.83}
\definecolor{c_polar}{rgb}{.98,0.93,1}
\definecolor{c_nonpolar}{rgb}{.87,0.87,.87}

\definecolor{c_SN}{rgb}{.4,0.,.4}
\definecolor{c_HT}{rgb}{.9,0.4,1}
\definecolor{c_VLIF}{rgb}{.5,0.5,.5}
\definecolor{c_C}{rgb}{1,0.6,0}
\definecolor{c_P}{rgb}{0,0.6,0}
\definecolor{c_G}{rgb}{1,0.6,0}
\definecolor{c_A}{rgb}{1,0.6,0}

\definecolor{notecolor}{rgb}{1,0.2,0.1}
\definecolor{linkcolor}{rgb}{0.2,0.2,1.0} 
\usepackage[pdftex,colorlinks=true, linkcolor= linkcolor, citecolor= linkcolor, urlcolor= linkcolor, hyperindex=true,hyperfigures=true]{hyperref} 

\usepackage{makecell}

\renewcommand{\vec}[1]{\mathbf{#1}}

\begin{document}

\title{Interpretable machine learning of amino acid patterns in proteins:
\\ a statistical  ensemble approach}

\author{Anna Braghetto}
\author{Enzo Orlandini}
\author{Marco Baiesi}
\address{Dipartimento di Fisica e Astronomia,
 Universit\`a di Padova, Via Marzolo 8, 35131, Padova, Italy
} 
\address{INFN, Sezione di Padova, Via Marzolo 8, 35131, Padova,
Italy}

\begin{abstract}
Explainable and interpretable unsupervised machine learning helps understand the underlying structure of data. 
We introduce an ensemble analysis of machine learning models to consolidate their interpretation. Its application shows that restricted Boltzmann machines compress consistently into a few bits the information stored in a sequence of five amino acids at the start or end of $\alpha$-helices or $\beta$-sheets.
The weights learned by the machines reveal unexpected properties of the amino acids and the secondary structure of proteins: (i) His and Thr have a negligible contribution to the amphiphilic pattern of $\alpha$-helices; (ii) there is a class of $\alpha$-helices particularly rich in Ala at their end; (iii) Pro occupies most often slots otherwise occupied by polar or charged amino acids, and its presence at the start of helices is relevant; (iv) Glu and especially Asp on one side, and Val, Leu, Iso, and Phe on the other, display the strongest tendency to mark amphiphilic patterns, i.e., extreme values of an {\em effective hydrophobicity}, though they are not the most powerful (non) hydrophobic amino acids.
\end{abstract}
\maketitle

\section{Introduction}
\label{sec:intro}

Various machine learning (ML) methods are applied to proteins~\cite{tubi19,wang2013predicting,karimi2019deepaffinity,karimi2020explainable,rodriguez2021feature,rube2022prediction,cai2022interpretable,ali2022interpretable,tubiana2022scannet,mata20,rives2021biological,alpha1,alpha2,baek21,hump21,drak22,torrisi2020deep,iuchi2021representation,wu2023survey,hermosilla2020intrinsic,wang2019high,ding2019deciphering,weigt2009identification,gligorijevic2021structure}.
For example, outstanding advancements have shown how ML can boost the prediction of protein native states~\cite{alpha1,alpha2,baek21} and complexes~\cite{hump21,drak22} based only on amino acid sequences. 
However, the aim of several approaches is not to achieve a reliable (black box) tool for protein structure prediction but to get informative knowledge from the big data available for protein sequences and structures. 

Interpretable ML~\cite{molnar2020interpretable,kamath2021explainable} focuses on understanding the cause of a model's decision and enhancing human capability to consistently predict the model's result. 
Interpretable ML versions are more complex and informative than standard statistical analysis and can improve our understanding of proteins~\cite{tubi19,wang2013predicting,karimi2019deepaffinity,karimi2020explainable,rodriguez2021feature,rube2022prediction,cai2022interpretable,ali2022interpretable,tubiana2022scannet}. In particular, they can detect patterns not emerging naturally from studying abundance and correlations of amino acids in secondary structures. Among the well-known patterns, for instance, there is the amphiphilic structure of several $\alpha$-helices and $\beta$-sheets~\cite{xion95,kamt99}, which are mostly (charged or) polar ($\mathbb P$) on one side and nonpolar ($\mathbb N$) on the other side. 
In an $\alpha$-helix, with pitch $\approx 3.6$ residues, the typical (non)polarity switch occurs every two residues.
On the other hand, in a $\beta$-sheet, the three-dimensional alternation of the side chains takes place at every step. Hence an amphiphilic sequence would be, for example, ${\mathbb{PNPNP}}$.

In this work, we use a simple form of interpretable unsupervised ML, restricted Boltzmann machines (RBMs)~\cite{smolensky1986information,hinton1986learning,hinton2012practical,tubi17,decelle2018thermodynamics,roussel2021barriers,fernandez2022disentangling,dece22,decelle2023unsupervised}, which allow extracting deep, nontrivial insight without losing the most transparent information on data statistics encoded in local biases. Conveniently, the weights and biases learned by RBMs can be visualized and easily interpreted. 
This established approach already revealed correlated amino acids within protein families~\cite{tubi19}, drug-target interactions~\cite{wang2013predicting}, and correlations within DNA sequences~\cite{si2016learning,di2022generative}.

A novelty of our work is a statistical ensemble approach to unsupervised ML, which improves the robustness of the findings. By training RBMs of the same size but with different weight initializations, we check whether they all converge to the same final set of learned weights. The maximally complex RBMs preserving this ensemble coherence are optimal, as they perform encoding of the correlations within data samples while providing stable and transparent information on the data. 

We show that our optimal RBMs perform extreme information compression to two or three bits, encoding the essential correlations between amino acids either at the beginning or at the end of $\alpha$-helices and $\beta$-sheets. 
In addition to recovering the expected amphiphilic structures, this approach (i) discovers more subtle yet relevant amino acid patterns in each portion of the secondary structure and (ii) provides evidence of similarity between amino acids' roles in these structures, including some surprising ones.

 RBMs distinguish two classes of amino acids, which we map to $\mathbb P$ and $\mathbb N$, as shown in Table~\ref{tab:1}. Contrary to standard classification~\cite{alberts2002molecular}, but similar to some partitionings (see references collected by \citet{step13}), Tyr belongs to the class $\mathbb N$ of hydrophobic amino acids. Pro is mostly $\mathbb P$, as discussed in detail below. Some surprising sub-classes emerge, especially by looking at the results in $\alpha$-helices, where it turns out that Thr and His play a similar weak role in the amphiphilic patterns. RBMs classify Asp and Glu on the one side and Val, Leu, Iso, and Phe on the other as the most diverse amino acids. However, Trp has the highest experimental hydrophobicity, while Arg and Lys have the lowest values~\cite{meil01}. To explain this finding, we argue that RBMs detect a kind of {\em effective hydrophobicity}, emphasizing how deeply amino acids play the hydrophobic or hydrophilic role in the amphiphilic alternation in $\alpha$-helices and $\beta$-sheets. These findings only partially overlap with those expressed by known diagrams of consensus amino acid similarity~\cite{step13}.

\begin{table}[!t]
 \centering
 \begin{tabular}{cllll}
 \toprule
 \multirow{9}{*}{$\mathbb P$}
 &\cellcolor{c_charged}&Arginine& Arg&\textcolor{red}{\bf \texttt{\ R}}\\
 &\cellcolor{c_charged}&Lysine& Lys&\textcolor{red}{\bf \texttt{\ K}}\\
 &\cellcolor{c_charged}&Histidine& His&\textcolor{c_HT}{\bf \texttt{\ \ \ H}}\\
 &\cellcolor{c_charged}&Aspartic acid& Asp&\textcolor{blue}{\bf \texttt{D}}\\
 &\multirow{-5}{*}{\cellcolor{c_charged}charged}
 &Glutamic acid& Glu&\textcolor{blue}{\bf \texttt{E}}\\
 &\cellcolor{c_polar} &Asparagine& Asn &\textcolor{c_SN}{\bf \texttt{\ \ N}}\\
 &\cellcolor{c_polar} &Glutamine& Gln &\textcolor{red}{\bf \texttt{\ Q}}\\
 &\cellcolor{c_polar} &Serine& Ser &\textcolor{c_SN}{\bf \texttt{\ \ S}}\\
 &\cellcolor{c_polar} &Threonine& Thr &\textcolor{c_HT}{\bf \texttt{\ \ \ T}}\\
 \hline
 \multirow{8}{*}{$\mathbb N$}
 &\multirow{-5}{*}{\cellcolor{c_polar} polar}
 &Tyrosine& Tyr&\textcolor{black}{\bf \texttt{Y}}\\
 &\cellcolor{c_nonpolar}&Tryptophan& Trp&\textcolor{black}{\bf \texttt{W}}\\
 &\cellcolor{c_nonpolar}&Valine& Val&\textcolor{c_VLIF}{\bf \texttt{\ V}}\\
 &\cellcolor{c_nonpolar}&Leucine& Leu&\textcolor{c_VLIF}{\bf \texttt{\ L}}\\
 &\cellcolor{c_nonpolar}&Isoleucine& Iso&\textcolor{c_VLIF}{\bf \texttt{\ I}}\\
 &\cellcolor{c_nonpolar}&Phenilalanine& Phe&\textcolor{c_VLIF}{\bf \texttt{\ F}}\\
 &\cellcolor{c_nonpolar}&Methionine& Met&\textcolor{black}{\bf \texttt{M}}\\\
 &\multirow{-7}{*}{nonpolar}\cellcolor{c_nonpolar}
 &Cysteine& Cys&\textcolor{black}{\bf \texttt{C}}\\
 \hline
 $\sim\mathbb P$
 &\cellcolor{c_nonpolar}&Proline& Pro&\textcolor{c_P}{\bf \texttt{\ P}}\\
 \hline
 &\cellcolor{c_nonpolar}&Glycine& Gly&\textcolor{c_G}{\bf \texttt{\ \ G}}\\
 \hline
  $\sim\mathbb N$&\cellcolor{c_nonpolar}&Alanine& Ala&\textcolor{c_A}{\bf \texttt{\ \ \ A}}\\
 \bottomrule
 \end{tabular}
 \caption{RBM-based classification of amino acids. The first column labels which amino acids can be classified as polar/hydrophilic ($\mathbb P$) and nonpolar/hydrophobic ($\mathbb N$), according to the weights of our RBMs. The second column shows the textbook classification of amino acids~\cite{alberts2002molecular}. According to RBMs, Tyr behaves as a nonpolar amino acid, Pro behaves mostly as a polar one ($\sim\mathbb P$), and Ala is slight $\mathbb N$ only in $\beta$-sheets. Gly is neither clearly $\mathbb P$ nor $\mathbb N$. The color of each amino acid symbol in the last column (and the offset) follows the sub-grouping we introduce based on the weights learned by the RBMs applied to $\alpha$-helices. }
 \label{tab:1}
\end{table}
\section{Methods}
\label{sec:met}

Here we describe the datasets and the methods used to analyze them. Some more technical details are reported in the Supplementary Information.

\subsection{Data}

To each protein sequence stored in the reduced CATH ensemble of natural proteins~\cite{CATH40}, we apply the algorithm DSSP~\cite{kabsch1983dictionary,joosten2010series} to determine the secondary structure to which every amino acid belongs. Then, we collect all sequences within $\alpha$-helices and $\beta$-sheets, long enough to contain $\Gamma=5$ amino acids. We then build four sets: one with the first $\Gamma=5$ amino acids in $\alpha$-helices (following the standard orientation from the N to the C terminus of the protein), one with the last $\Gamma$ amino acids in $\alpha$-helices, and the same for two more sets at the start and end of $\beta$-sheets. The two sets referring to the first and last $\Gamma=5$ amino acids in $\alpha$-helices contain $129300$ samples each, while the other two sets, concerning the start and end of $\beta$-sheets, include $101382$ sequences each.

We use a one-hot encoding to represent amino acids. Namely, the $k$-th amino acid is stored as a sequence $\vec v^k = (-1,-1,\ldots,+1,\ldots,-1,-1)$ of $20$ integers where only a $+1$ element is present at position $k$.
This encoding is how an RBM reads the amino acid in a portion of its visible units. A sequence of $\Gamma$ amino acids $(k_1,\ldots,k_\Gamma)$ is thus translated into one-hot encoding stacked as $\vec v = (\vec v_1^{k_1},\vec v_2^{k_2},\ldots\vec v_\Gamma^{k_\Gamma})$ giving a total of $N_v=20\cdot\Gamma$ digits in a data sample.

To monitor the training of each RBM, we randomly split data into a training (80\%) and a validation (20\%) set. The training set is used to optimize the RBM parameters and compute the pseudo log-likelihood (PLL) function, which measures the quality of data reconstruction by RBMs~\cite{goodfellow2016deep}. The PLL of the validation set is then used to check the performance of the RBM in reproducing the statistics of new data. Note that, in principle, this procedure could cause differences in the results. However, the cases we find in the ensemble of RBMs, as explained below, reveal when variability is small and highlight general patterns.

\begin{figure}[t!]
 \centering
 \includegraphics[width=0.45\textwidth]{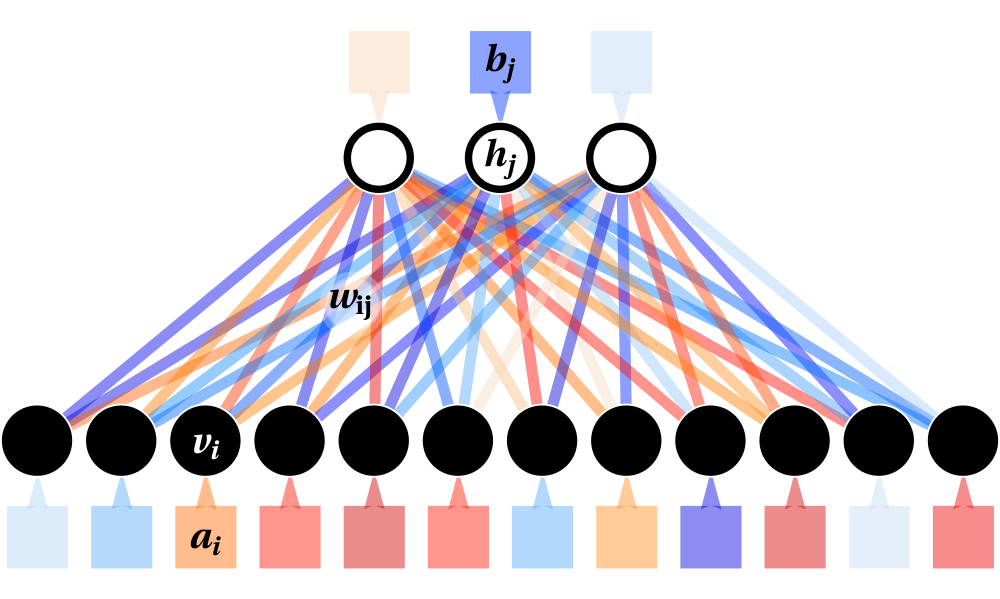}
 \caption{Sketch of an RBM with $N_v=12$ visible units (black circles, where data are given as an input) and $N_h=3$ hidden units (white circles). Red and blue shades indicate positive and negative values of single weights (plotted as lines joining units in the two layers) and biases (boxes next to units). We will follow a similar color scheme in Figures 4-7 below.} 
 \label{fig:sketch}
\end{figure} 

\subsection{Restricted Boltzmann Machines}

The RBM is an unsupervised machine learning method based on a simple neural network architecture. It aims to reproduce the empirical distribution of data samples by encoding the correlations between their elements, the {\em visible units} $v_i$ ($1\le i \le N_v$). This encoding uses a set of parameters and a layer of hidden (or latent) variables $h_j$ ($1\le j\le N_h$).
The parameters defining the method are the weights $w_{ij}$ in a $N_v\times N_h$ matrix connecting visible to hidden units and the local biases that act both on the visible ($a_i$) and the hidden ($b_j$) units.
Figure~\ref{fig:sketch} shows a sketch of an RBM. 
The statistical weight of a $(\vec v, \vec h)$ configuration is given by
\begin{align}
 e^{-E(\vec v, \vec h)} =
 \exp\bigg(
 & \sum_{i=1}^{N_v}\sum_{j=1}^{N_h} v_i w_{ij} h_j \nonumber\\
 & +\sum_{i=1}^{N_v} a_i v_i +
\sum_{j=1}^{N_h} b_j h_j\bigg)
\label{we}
\end{align}
It resembles a Boltzmann weight with energy $E(\vec v, \vec h)$, for which we will use "spin" variables $v_i=\pm 1$, $h_j=\pm 1$. Since this version generates a finite number $2^{N_h}$ of hidden states, it facilitates an interpretation of the structure of weights between hidden and visible units and of local biases. Initially, weights and biases of untrained RBMs are drawn randomly from chosen distributions.

The bipartite structure of the RBM allows an easy generation of $\vec h$ from $\vec v$. This step should encode the correlations within data sequences in $N_h$ hidden units for a trained RBM. When $N_h\ll {N_v}$ the RBM acts as an {\em information bottleneck} enforcing 
such a simple model, with its small resources, to capture the crucial properties of the analyzed data.
The $\vec v \to \vec h$ step selects each $h_i$ independently with probability
\begin{equation}
\label{forw}
 p(h_j|\,\vec{v}) \sim
 \exp\left[h_j \left(b_j + \sum_{i=1}^{N_v} w_{ij}v_i \right)\right]
\end{equation}

Similarly, one generates
$\vec v$ if $\vec h$ is known, through $\vec{v} \sim p(\vec{v}|\vec{h})$.
Each of the $\Gamma$ blocks $\vec v_\gamma$ is generated independently.
The indices $i\in I(\gamma)$ of the 20 weights $w_{ij}$ pointing to the segment $\vec v_\gamma$ are those relevant for its sampling. By remapping these indices $i$ to the interval $k=1,...,20$, we pick an amino acid $k$ with probability
\begin{align}
 \label{back}
 p(k\,|\vec h) &\sim e^{2\, \varphi_k (\vec h)}
 \\
 \text{with}\quad \varphi_k(\vec h)
 & = a_k + \sum_{j=1}^{N_h} w_{kj} h_j\;,
 \label{phi}
\end{align}
where $\varphi_k(\vec h)$ is the local field on the site $k$.

The core of the training of an RBM consists of sampling values in visible and hidden units through an algorithm termed contrastive divergence with $n$ Monte Carlo steps (CD-$n$)~\cite{hinton2002training,mehta2019high}. 
 We alternatively sample from conditional distributions, starting from a data sample $\vec v_0$ at $t=0$ up to $t=n$ steps: $\mathbf{h}_{t+1} \sim p(\mathbf{h}|\mathbf{v}_t)$ and $\mathbf{v}_{t+1} \sim p(\mathbf{v}|\mathbf{h}_{t+1})$.
The statistics of the sampled configurations allow the estimation of the gradient of the data log-likelihood according to the Boltzmann weight ~\eqref{we}, in the direction of all parameters $w_{ij}$, $a_i$, and $b_j$. Then, we apply a standard gradient ascent algorithm (in our case, Adam~\cite{mehta2019high}).
In addition to CD-$n$, we use persistent CD-$n$ (PCD-$n$), a variant that should better sample the configurational space~\cite{tieleman2008training,mehta2019high}.

Once trained, the same sampling procedure may generate realistic amino acid sequences $(k_1,\ldots,k_\Gamma)$. Through step \eqref{back}, an RBM decides how to decode the hidden units $\vec h$ and generate a sequence. 

\subsection{Ensemble of RBMs}

A novelty of this work is using a statistical ensemble analysis of machine learning. To this purpose, we run $R$ independent realizations of RBMs with the same size ${N_v}, N_h$, and with the same training epochs (we found that typically 50 or 100 epochs are enough to train the model). 
Note that each realization differs by the weights initialization and the random splitting of the data in training and validation sets.

\begin{figure}[t!]
 \centering
 \includegraphics[width=0.48\textwidth]{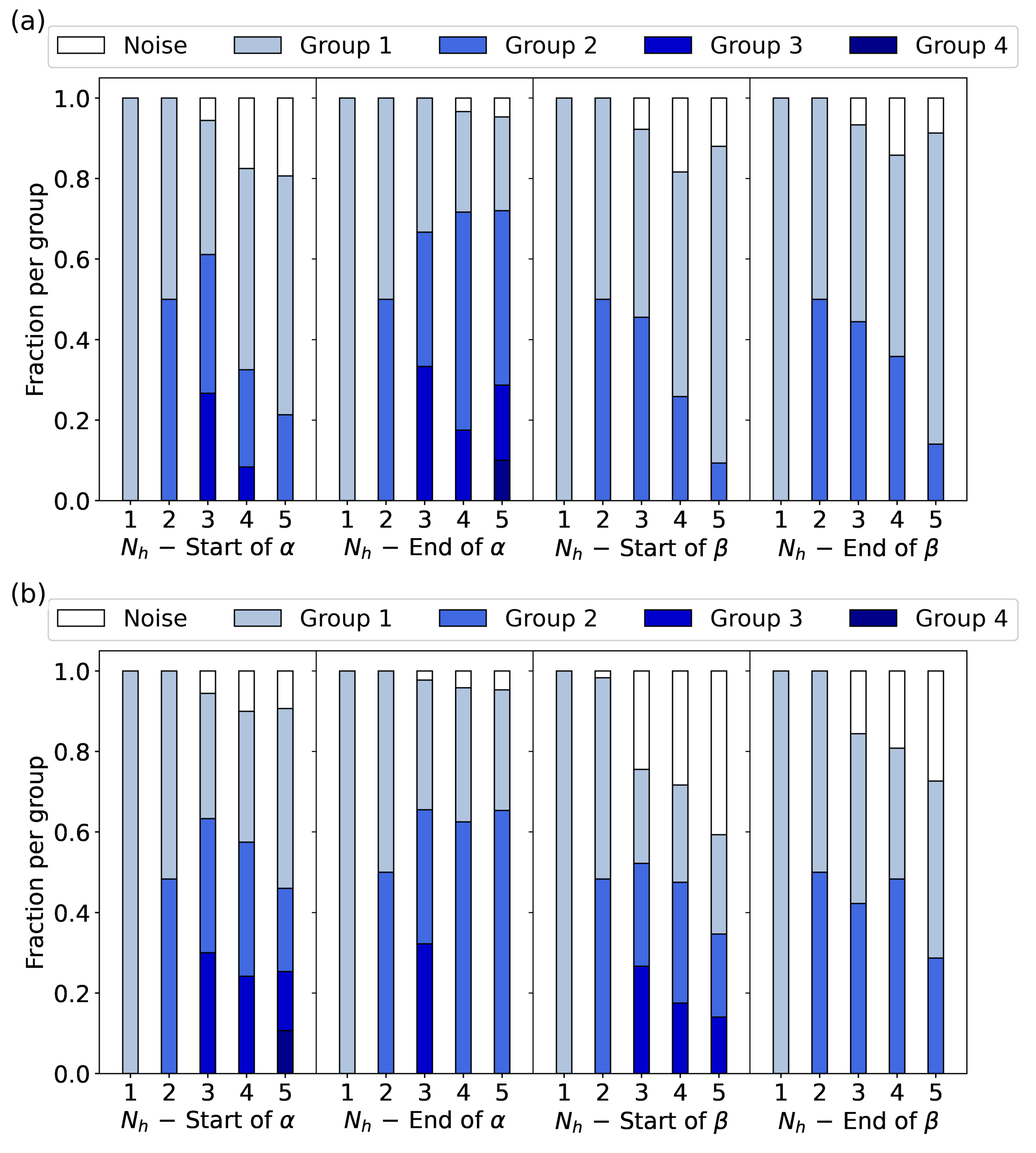}
 \caption{For (a) CD-1 and (b) PCD-10, we show the number and relative size of groups emerging from clustering hidden units in the ensemble of RBMs for every position of the secondary structure that we study. In both cases, we conclude the following: for $\alpha$-helices, $N_h=3$ is the optimal number of hidden units, while for $\beta$-sheets, it is $N_h=2$. These are the maximum values where the number of groups matches the number of hidden units, and the noise is still tiny, i.e., where each RBM in the ensemble has learned the same set of hidden units.}
\label{fig:groups}
\end{figure} 

\begin{figure*}[t!]
 \centering
 \includegraphics[width=.85\textwidth]{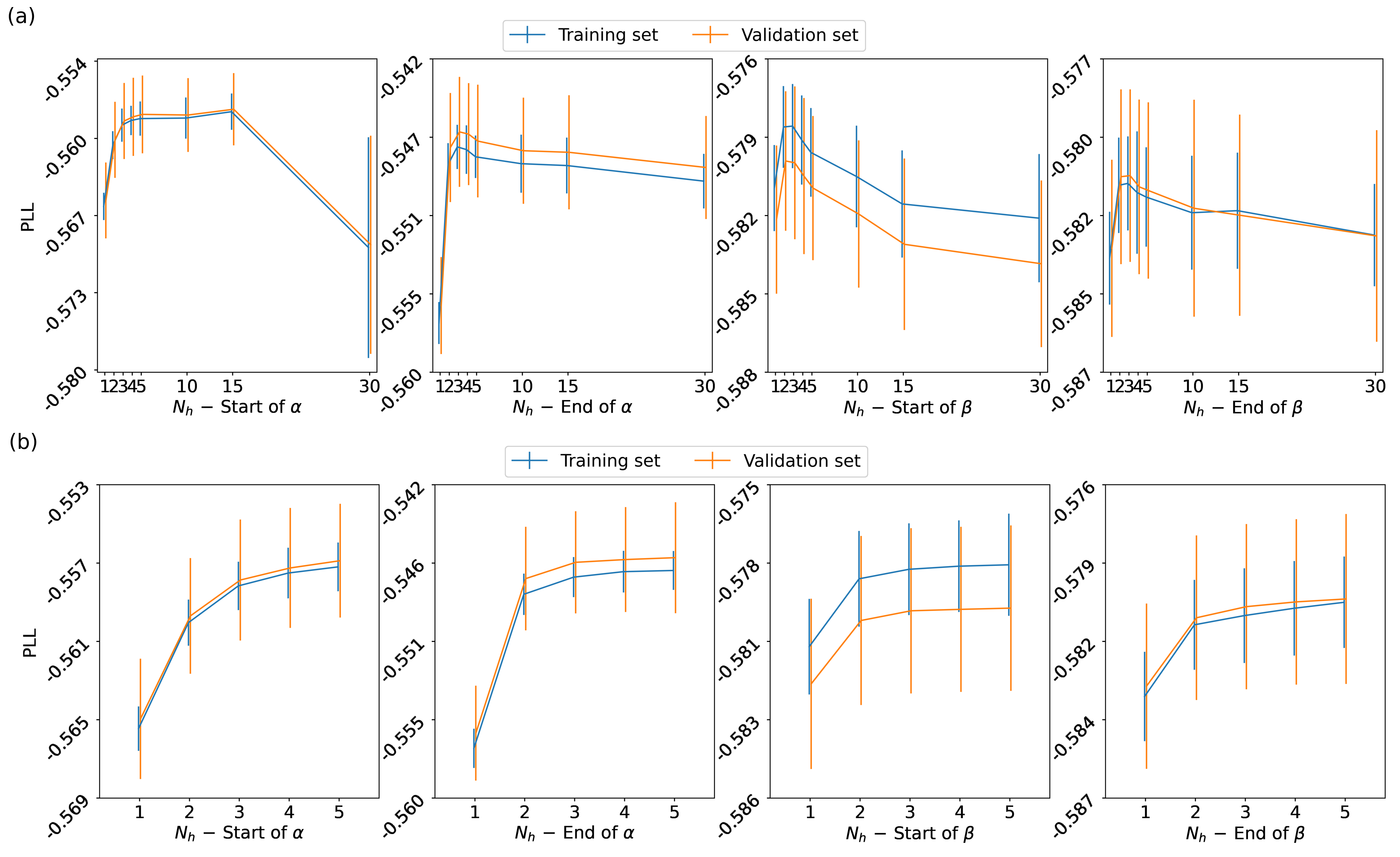}
 \caption{Pseudo log-likelihood as a function of the number of hidden units for RBMs trained with (a) CD-1 and (b) with PCD-10, shown for each of the four segments of secondary structure that we study. The PLL for the train and the validation set are compatible, showing that the RBMs have achieved robust training.} 
 \label{fig:pseudo}
\end{figure*} 

From now on, let us assume that each hidden unit $j$ in a single RBM is characterized by its set of weights $w_{ij}$ ($1\le i\le {N_v}$). All hidden units from RBM realizations are then collected in an ensemble of hidden units to study their properties and check if common patterns exist. 
For every pair of hidden units $j,m$, the Euclidean distance $d_{jm} = \sqrt{\sum_i |w_{ij} - w_{im}|^2}$ estimates their similarity. When used in a clustering algorithm, it identifies {\em groups} of hidden units (see the Supplementary Information for more details).

By averaging the weights $w_{ij}$ within each group and the biases $a_i$ and $b_j$ of each RBM, we build their average RBM (aRBM): this is supposed to represent the best summary of the relevant information learned by the ensemble. First, we use the aRBM to compute the probability \eqref{forw} of the $2^{N_h}$ possible hidden states, given that $\vec v$ are all points in a dataset.
Then, from hidden states weighted with their probability, we use \eqref{back} to verify the ability of the aRBM to faithfully reproduce the statistics of the original dataset in the visible space.

\section{Results}
\label{sec:res}

\subsection{Selecting the number of hidden units}
\label{sec:res1}
The main aim is to find simple patterns representative of the redundant, generic correlations in amino acid sequences (at the start of $\alpha$-helices, etc.) while neglecting specific patterns of single sequences with RBMs. The key to achieving this goal is the information bottleneck obtained by setting a small number $N_h$ of hidden units.

We monitor how many groups of hidden units emerge by increasing $N_h$ (Figure~\ref{fig:groups}a for CD-1 training of RBMs and Figure~\ref{fig:groups}b for PCD-10). Generally, the ratio of groups to hidden units stays maximal up to $N_h=3$ for $\alpha$-helices and $N_h=2$ for $\beta$-sheets. For these values of $N_h$, almost all RBM realizations have the same palette of hidden units. Only in a few cases, the clustering algorithm (see the Supplementary Information) classifies units as {\em noise} due to their significant diversity from all other ones. Note that beyond these values of $N_h$, there is no clear one-to-one correspondence between hidden units in an RBM and groups, and uniformity in the ensemble of RBMs is lost.

To evaluate the performance of the RBMs, we compute the PLL as a function of $N_h$, see Figure~\ref{fig:pseudo}. From $N_h=1$, for CD-1 and PCD-10, the PLL quickly reaches a plateau around $N_h\approx 3$. By adding more hidden units ($N_h > 3$), one does not obtain any significant improvement in the performance of the RBM. Moreover, for CD-1, we can go up to $N_h=30$, finding in all cases a decreasing trend of the PLL for large $N_h$. Hence, more complex RBMs are heterogeneous and suboptimally trained.

All considered, we show the results for $N_h=3$ for $\alpha$-helices and $N_h=2$ for $\beta$-sheets.
From now on, we will discuss only the results from PCD-10. Those from CD-1 are similar.

\begin{figure*}[t!]
 \centering
 \includegraphics[width=0.95\textwidth]{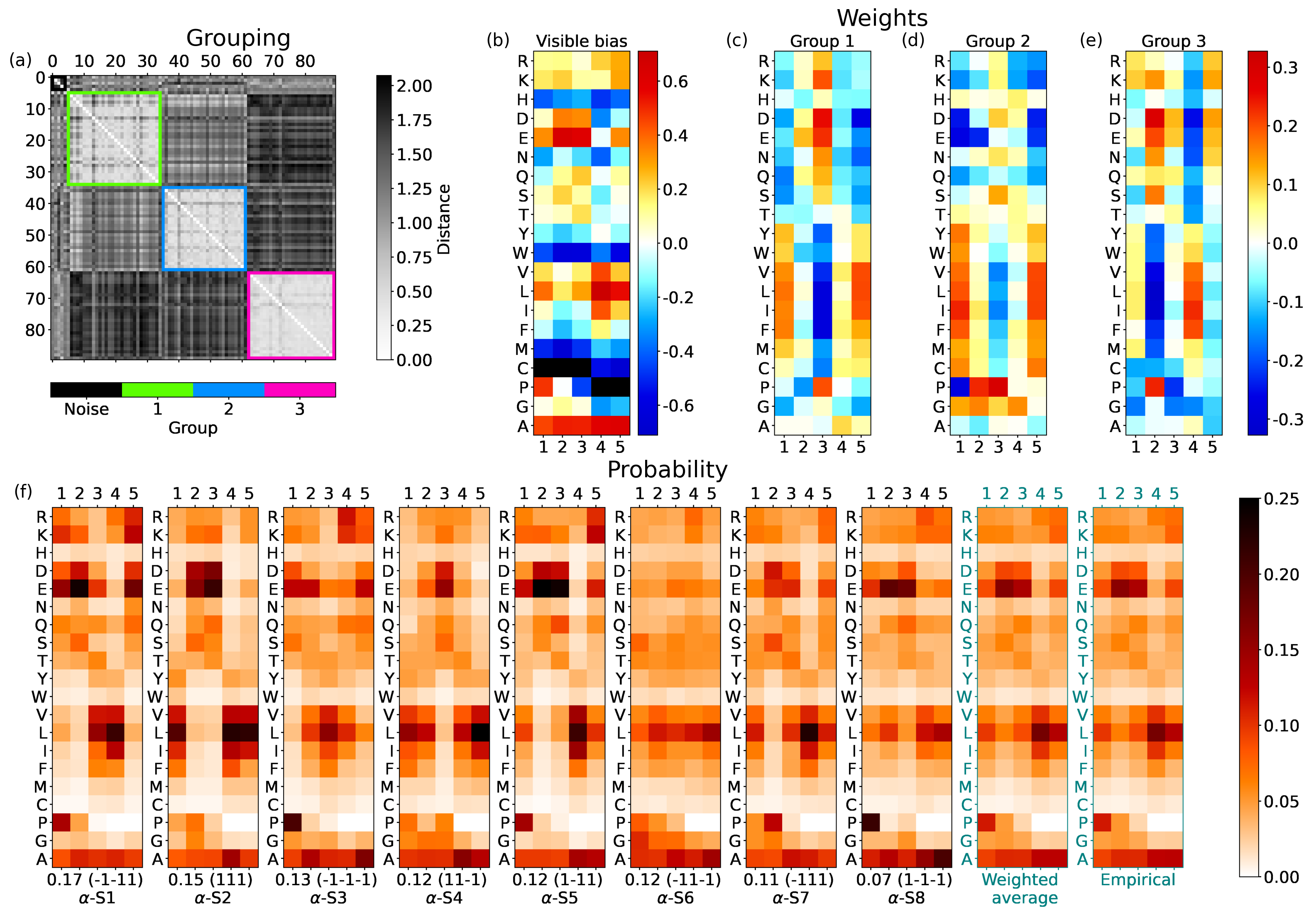}
 \caption{For the start of $\alpha$-helices, with ${N_h}=3$ hidden units: (a) Matrix with gray shade indicating the distance $d_{jm}$ between the weights $w_{ij}$ and $w_{im}$ of different hidden units $j,m$; the color boxes highlight the groups found by the DBSCAN clustering. (b) Average biases $a_i$ learned by the ensemble of RBMs, reshaped from an array with $20\Gamma=100$ entries to a $20\times \Gamma$ table, in which each column corresponds to a given encoding $\vec v_\gamma$ and each row to a given amino acid (a similar scheme is used in panels (c), (d), (e)). Values more negative than the lower threshold in the scale are marked with black squares (in this case for Cys and Pro, which essentially leads to the negligible probability of finding these amino acids in those positions). (c), (d), and (e): Average weights of units in groups 1, 2, and 3. 
 (f) The shade of each slot in each panel shows the probability of picking a specific amino acid at a given position. Hence columns are normalized to $1$. The first $2^{N_h}=8$ panels show the probabilities for every hidden state (the sequence of $\pm 1$'s in the parenthesis at the bottom, where it follows the value of its empirical frequency).
 Hidden states are labeled and ranked with decreasing frequency, e.g., $\alpha$-S1 is the most probable hidden state at the start of $\alpha$-helices.
 The last two panels show the average of RBM $\alpha$-S states weighted according to their frequency, and the actual probability of amino acids at the $\Gamma=5$ initial positions of $\alpha$-helices. In practice, the prescription of the RBM for reconstructing meaningful sequences would be to (i) pick a hidden state at random according to its frequency and (ii) according to probabilities in its table, for every position $\gamma \le \Gamma$ pick an amino acid at random.
 The values of the hidden bias in the aRBM for each group are $b_1=-1.129$, $b_2=1.270$, $b_3=-1.496$.}
 \label{fig:a1}
\end{figure*} 

\subsection{How to read weight patterns}
\label{sec:res2}
We summarize the properties of the RBMs' ensemble via a set of plots, as reported, for instance, in Figure~\ref{fig:a1} for the starting strand of $\alpha$-helices.
We average the values for weights in a group or biases from all RMBs in the ensemble. Thus, the displayed values represent the aRBM.

Figure~\ref{fig:a1}(a) reports the table of distances between any two hidden units of the ensemble of RBMs. The units are sorted and collected into the groups (colored squares with light internal colors along the diagonal) detected by the clustering algorithm. Some units, marked as "noise", are not assigned to any group.

Figure~\ref{fig:a1}(b) shows the visible bias $a_i$ of the aRBM, reshaped to a $20\times \Gamma$ matrix for better readability. The exponential of this bias is a good indicator of the mean probability of finding an amino acid at a specific position in the sequence. For instance, it shows that Glu (E) has a high chance of appearing at positions 2 and 3. 
For each of the groups, we show in Figure~\ref{fig:a1}(c), (d), and (e) the weights $w_{ij}$ of the corresponding hidden unit in the aRBM. In addition, the $N_h$ biases $b_j$ of the aRBM are specified in the caption.

A table as the one reported in Figure~\ref{fig:a1}(c) may be read as follows: a hidden unit in group 1 "pushes'' a random pattern of amino acids biased by the weights toward a particular sequence, depending on its stored value $h_1=\pm 1$.
According to \eqref{back} and \eqref{phi}, $h_1= 1$ raises the probability of picking amino acids with weights $w_{i1}$ to a value significantly larger than zero (red shades). For instance, one can notice that D is chosen more frequently at position 3 and I, L, V, and F at positions 1 and 5. The opposite happens if $h_1=-1$.

All other random choices are possible with gradually lower probability.
The unit does not (de)select any particular amino acid when weights have values close to zero (light colors in the table). Instead, another unit may be the one that drives the sequence selection in that position. For instance, in Figure~\ref{fig:a1}(c)-(d), we see that units in group 1 and group 2 have a strong set of weights at positions 1, 3, and 5, which are complementary to those of group 3 (stronger at positions 2 and 4, see Figure~\ref{fig:a1}(e)). Therefore, the units in different groups may take care of different alternating slots in the sequence.

The first $2^{N_h}=8$ panels in Figure~\ref{fig:a1}(f) represent probabilities \eqref{back} to choose amino acids at every slot $\gamma \le \Gamma$ (normalized in columns at fixed $\gamma$) if the aRBM is in a given state $\vec h$. 
With \eqref{forw}, we compute the probability of each of the $2^{N_h}=8$ hidden states $\vec h$ from biases $b_j$, weights $w_{ij}$, and $\vec v$ in the dataset. This is indicated below each panel in Figure~\ref{fig:a1}(f) (e.g.,~$0.18$). After this, we also specify the state $\vec h$ (e.g.,~$(1,-1,-1)$) and a chosen label (e.g.,~$\alpha$-S1).
We rank the $\vec h$ states in Figure~\ref{fig:a1}(f) from the most frequent to the least likely. 

Each of the first $2^{N_h}$ panels of Figure~\ref{fig:a1}(f) thus displays a typical correlation of probabilities followed by the aRBM to build a sequence of amino acids.
The last two panels are the average of the first $2^{N_h}$ panels, weighted with their frequency, and the empirical average in the dataset.

In the following, we specify the discussion of the four regions of secondary structure analyzed in this work.

\subsection{Start of $\alpha$-helices}
\label{sec:a1}
The first training set we study contains stretches of the first $\Gamma=5$ positions in all (long enough) $\alpha$-helices in proteins of the CATH database.
The corresponding set of trained RBMs with ${N_h}=3$ yields three significant groups of hidden units, see Figure~\ref{fig:a1}(a).
For ${N_h}=1,2,3$, we see appearing, respectively, groups 1, 3, and finally 2. 
By including additional hidden units, we continue to observe these three groups, confirming that RBMs encode the main patterns within the analyzed sequences with three hidden units.

Figure~\ref{fig:a1}(b) shows the bias $a_i$ of the aRBM. It is quite structured compared to other cases, shown later, as the end of $\alpha$-helices and $\beta$-sheets. This structure denotes a tendency of amino acids to appear more frequently at specific positions.
Notice the pattern of Pro, with high intensity (red) at the first position, which sensibly decreases in the next positions (the black color means that $a_i$ is below the lower level of the scale), in agreement with the known abundance of Pro at the start of helices~\cite{kim1999positional}.
Notably, at position $\gamma=4$, there stands out a peculiar behavior: a high intensity for nonpolar amino acids (in particular Val (V), Leu (L), and Iso (I)) aligns with a low intensity for polar amino acids (especially Asp (D), Glu (E) and Asn (N)). Consistently, an average depletion of polar amino acid at position $\gamma=4$ at the start of $\alpha$-helices is visible in the empirical statistics, shown in the last panel of Figure~\ref{fig:a1}(f).

In addition to the average trend dictated by the bias, the aRBM, thanks to the hidden units, can modulate the correlations among amino acids in single sequences.
Hidden units in group 1 (Figure~\ref{fig:a1}(c)) address anticorrelations between $\mathbb P$ and $\mathbb N$ amino acids at positions $\gamma=1,3,5$. For instance, $h_1=1$ promotes the pattern $\mathbb N$-$\mathbb P$-$\mathbb N$ while $h_1=-1$ promotes $\mathbb P$-$\mathbb N$-$\mathbb P$. 
Group 3 (Figure~\ref{fig:a1}(e)) instead mainly encodes the correlations among amino acids at positions $\gamma=2,4$. 
Group 2 (Figure~\ref{fig:a1}(d)) is similar to group 1 but also displays a set of large weights for Pro. This set adds significant insight to the correlations between Pro as a starter of helices and its following amino acids (the bias did not show such a rich structure): for example, weights in group 2 suggest that P, E, and D are interchangeable at the position $\gamma=1$ and that they are strongly correlated with D, E at $\gamma=5$ and anticorrelated with P at $\gamma=3$.

Given the aRBM, we check the states in the hidden space in Figure~\ref{fig:a1}(f), allowing us to merge the information from biases and weights. Different configurations appear, but almost all show a repeated scheme with polar and nonpolar amino acids alternation with blocks of about two elements, consistent with an amphiphilic structure in $\alpha$-helices. More interestingly, states $\alpha-$S1, $\alpha-$S3, $\alpha-$S5, $\alpha-$S8 (sharing a $h_2=-1$ that promotes Pro in group 2) include the activation of Pro at the start of the sequence, paired with Glu in the second position (for this subset of $\alpha$-helices, we notice that Glu's activation is not fixed only at the second position but is active also at the first or third position). 
This pattern provides two main classes of amino acid alternation: (Pro)$\mathbb{PNNP}$ or (Pro)$\mathbb{PPNN}$. In this context, Pro behaves as polar, with higher frequency ($\alpha-$S1, $\alpha-$S3), or as nonpolar, with lower frequency ($\alpha-$S5, $\alpha-$S8).

We have thus shown that training led the RBMs to automatically detect and decompose the start of $\alpha$-helices into eight nontrivial modes. The reverse, trivial process of averaging their probabilities leads to the average behavior shown in the second-to-last panel of Figure~\ref{fig:a1}(f), which matches the empirical probabilities (last panel). Notably, the RBM decomposition would not be accessible, a priory, by standard statistical tools. Moreover, the discovered heterogeneous eight modes {\em generate} synthetic sequences, each with its own probabilistic pattern.

\begin{figure*}[tbh!]
 \centering
 \includegraphics[width=0.95\textwidth]{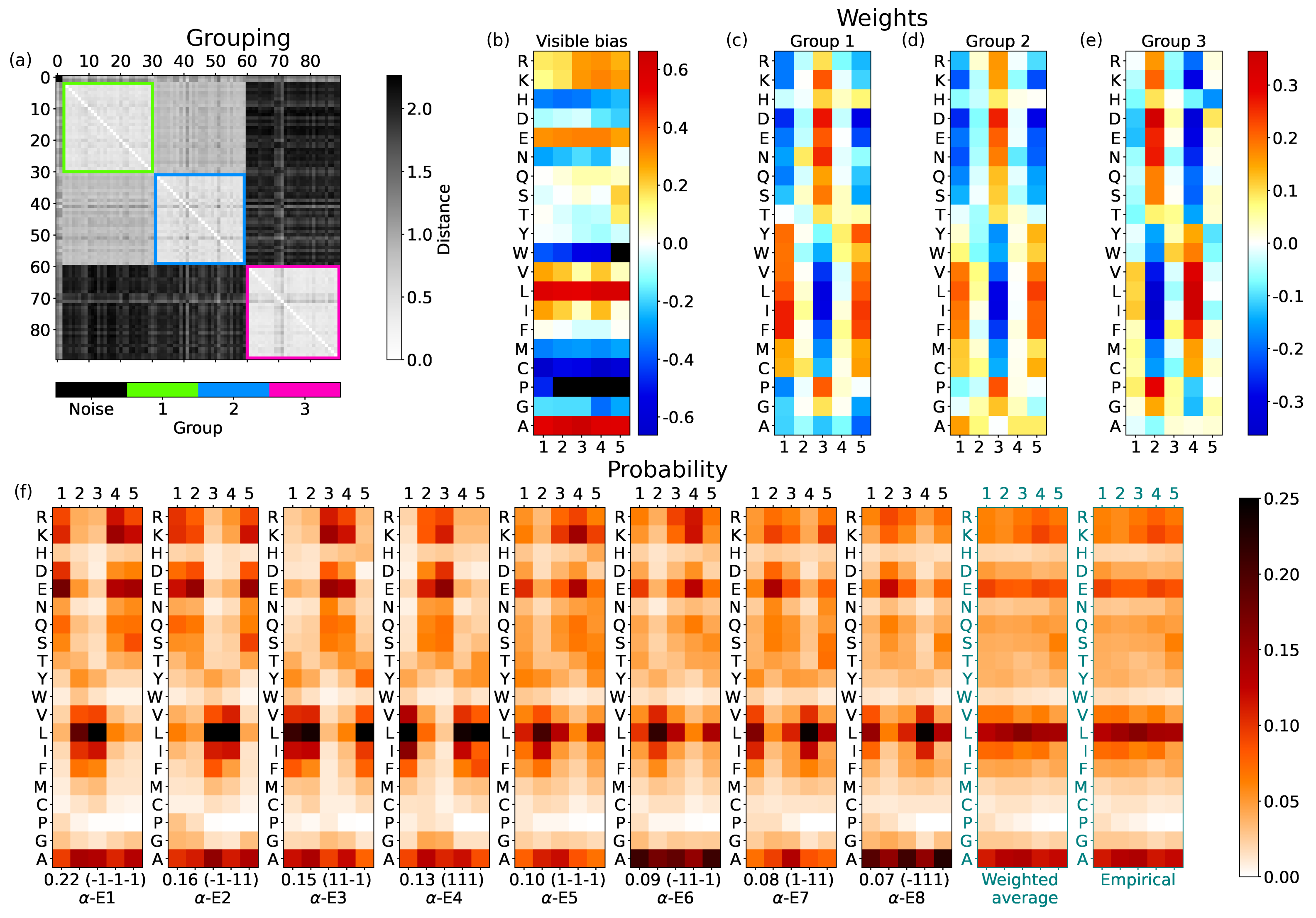}
 \caption{For the end of $\alpha$-helices with 3 hidden units, the same scheme as in Fig.~\ref{fig:a1}.  Hidden bias: $b_1=0.541$, $b_2=-0.410$, $b_3=-0.191$.}
 \label{fig:a2}
\end{figure*} 

\subsection{End of $\alpha$-helices}
\label{sec:a2}

The results from RBMs with ${N_h}=3$ for the last $\Gamma=5$ amino acids of $\alpha$-helices are displayed in Figure~\ref{fig:a2}. 
Again, three groups of hidden units emerge from clustering their weights. For ${N_h}=4$, they would remain the same. However, by increasing $N_h$ from $N_h=1$, we note that groups 1 and 2 are represented by their averaged version for $N_h\le 2$, while they split for $N_h=3$. This splitting is convincing: indeed, the PLL slightly increases in the $N_h=2\to 3$ step (Figure~\ref{fig:pseudo}) and, above all, the division into separate groups by the clustering is clear, see Figure~\ref{fig:a2}(a). 

Groups 1 and 2 determine the alternation of $\mathbb P$ and $\mathbb N$ at positions $\gamma=1,3,5$.
What distinguishes them is the weight pattern of Ala, which flips its sign from one group to the other, see Figure~\ref{fig:a2}(c) and (d). 
Group 3 instead fixes the alternation of $\mathbb P$ and $\mathbb N$ at positions $\gamma=2,4$.

The visible bias in Figure \ref{fig:a2}(b) shows that amino acids distribute almost uniformly at different positions at the end of $\alpha$-helices, with a significantly high bias toward Leu (L) and Ala (A). However, some slight deviations from the general behavior are visible.
For example, at the last position of the helix ($\gamma=5$), some polar amino acids are more probable (see N, S, and T), while some nonpolar ones become less likely (see V, I).
Note also the low bias of Gly at the last but one position $\gamma=4$.

In the hidden space, the aRBM reproduces, on average, the visible statistics, see Figure~\ref{fig:a2}(f). As observed at the start of $\alpha$-helices, some states ($\alpha$-E1,$\alpha$-E2,$\alpha$-E3,$\alpha$-E4) report the polar and nonpolar alternation with period $\sim 2$. More interestingly, in the states $\alpha$-E6 and $\alpha$-E8, the behavior of Ala spikes with a high probability in every position.
As known, Ala is a strong helix stabilizer~\cite{zhuang2021energetics,mier2022sequence}. Consistently, Ala has a high bias $a_i$ in the RBM and thus can act as a wild card: its placement in a typical sequence ending an $\alpha$-helix is relatively free, and it fits even at the specific positions of charged and polar amino acids. This high bias was also visible at the start of $\alpha$-helices (Figure~\ref{fig:a1}(b)), where no weight pattern induces the splitting of hidden units into separate groups based on Ala. The boosted probability of Ala in states $\alpha$-E6 and $\alpha$-E8 reveals a subclass of $\alpha$-helix endings ($15\%$ of the cases) richer in Ala than typical $\alpha$-helices. We have verified a posteriori that AAAAA is among the ten most frequent sequences at the end of $\alpha$-helices. Hence, poly-alanine~\cite{mier2022sequence} is a characterizing feature of the terminal part of $\alpha$-helices.

\begin{figure*}[t!]
 \centering
 \includegraphics[width=0.8\textwidth]{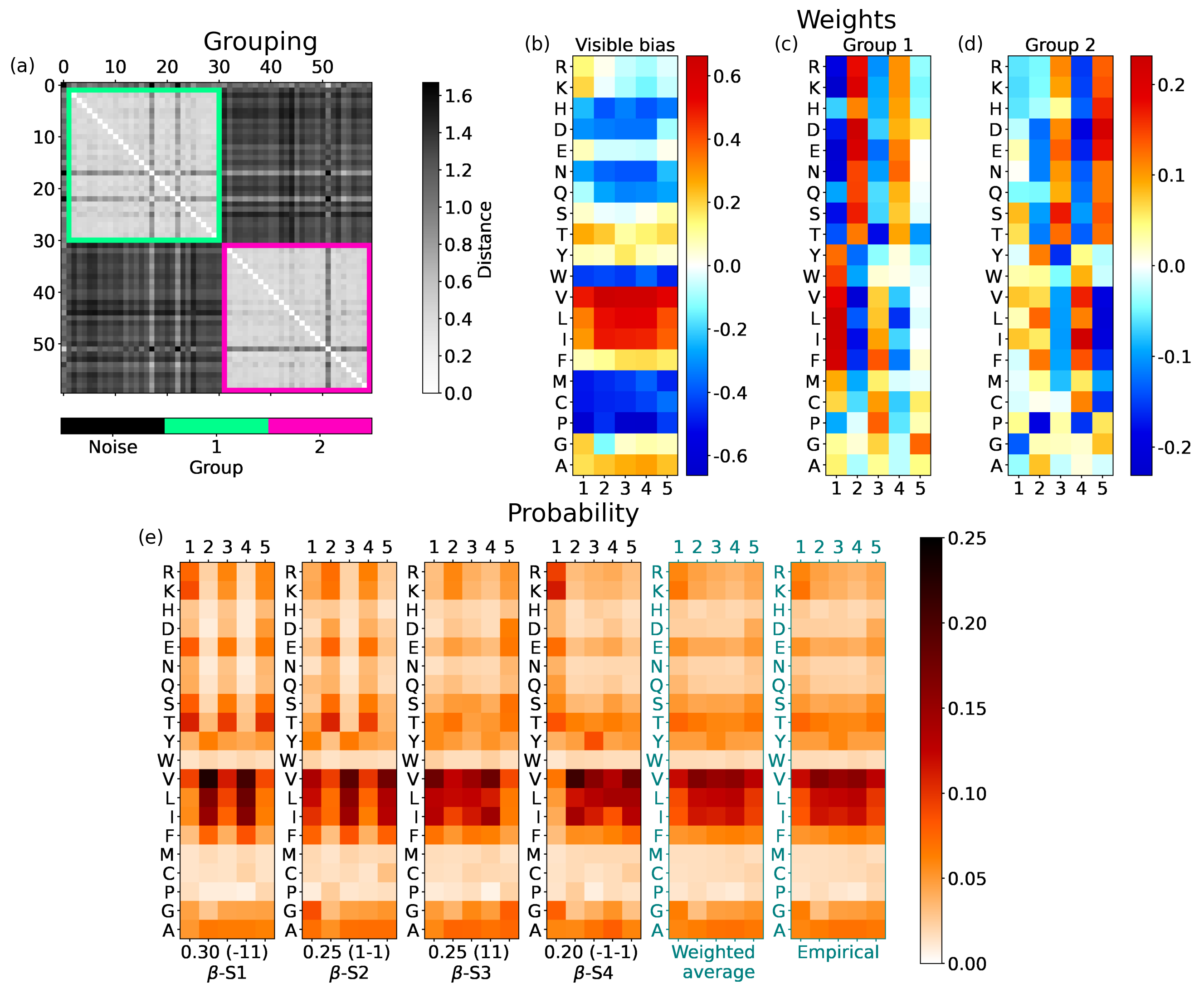}
 \caption{ For the start of $\beta$-sheets with 2 hidden units, the same scheme as in Fig.~\ref{fig:a1}. 
 Hidden bias: $b_1=0.458$, $b_2=-0.002$.}
 \label{fig:b1}
\end{figure*} 

\subsection{Start of $\beta$-sheets}
\label{sec:b1}

An alternating sequence $\mathbb{PNPN\ldots}$ of polar and nonpolar amino acids may allow $\beta$-sheets to expose side chains of the same kind at each of their two sides, making them amphiphilic.
For $N_h=1$, we find that the single hidden unit has weights of alternating signs with $\gamma$ and opposite polarity for $\mathbb P$ and $\mathbb N$, which would often lead to generating amphiphilic sequences. 
However, not all $\beta$-sheet stretches follow this simple amphiphilic scheme. For $N_h=2$, two groups emerge from clustering. The three hidden unit groups emerging for $N_h=3$ instead invalidate the analysis based on the aRBM for two reasons. First, many units are considered noise by the clustering algorithm; second, within single RBMs, we find high heterogeneity in the combination of groups.
Therefore, we choose $N_h=2$ as the optimal number of hidden units leading to the most consistent yet complex aRBM.
In support of this choice, note (Figure~\ref{fig:pseudo}) that the most significant increase in the PLL occurs from $N_h=1$ to $N_h=2$.

The weights of the two groups preserve the $\mathbb P\mathbb N$ alternation only at the beginning (group 1, Figure~\ref{fig:b1}(c)) or at the end (group 2, Figure~\ref{fig:b1}(d)). These will yield a hidden state $\vec h$ compatible with the amphiphilic pattern of weights if combined with the proper signs of $h_1$ and $h_2$: the probability of amino acids, for mode $\beta$-S1 (Figure~\ref{fig:b1}(e)), promotes the $\mathbb{PNPNP}$ alternation, while for mode $\beta$-S2 promotes the $\mathbb{NPNPN}$ pattern. They cover $55\%$ of the cases. 

However, the remaining $45\%$ of combinations of hidden states suppress the $\mathbb{PN}$ alternation and $\beta$ segments $\mathbb{NNNNP}$ (mode $\beta$-S3) and $\mathbb{PNNNN}$ ($\beta$-S4) are more likely to be generated by RBMs. In particular, $\beta$-S4 shows a strong activation of polar amino acids in the first position of $\beta$-sheets, in comparison to the aliphatic ones, which are instead very favored in the next four positions.

\begin{figure*}[t!]
 \centering
 \includegraphics[width=0.8\textwidth]{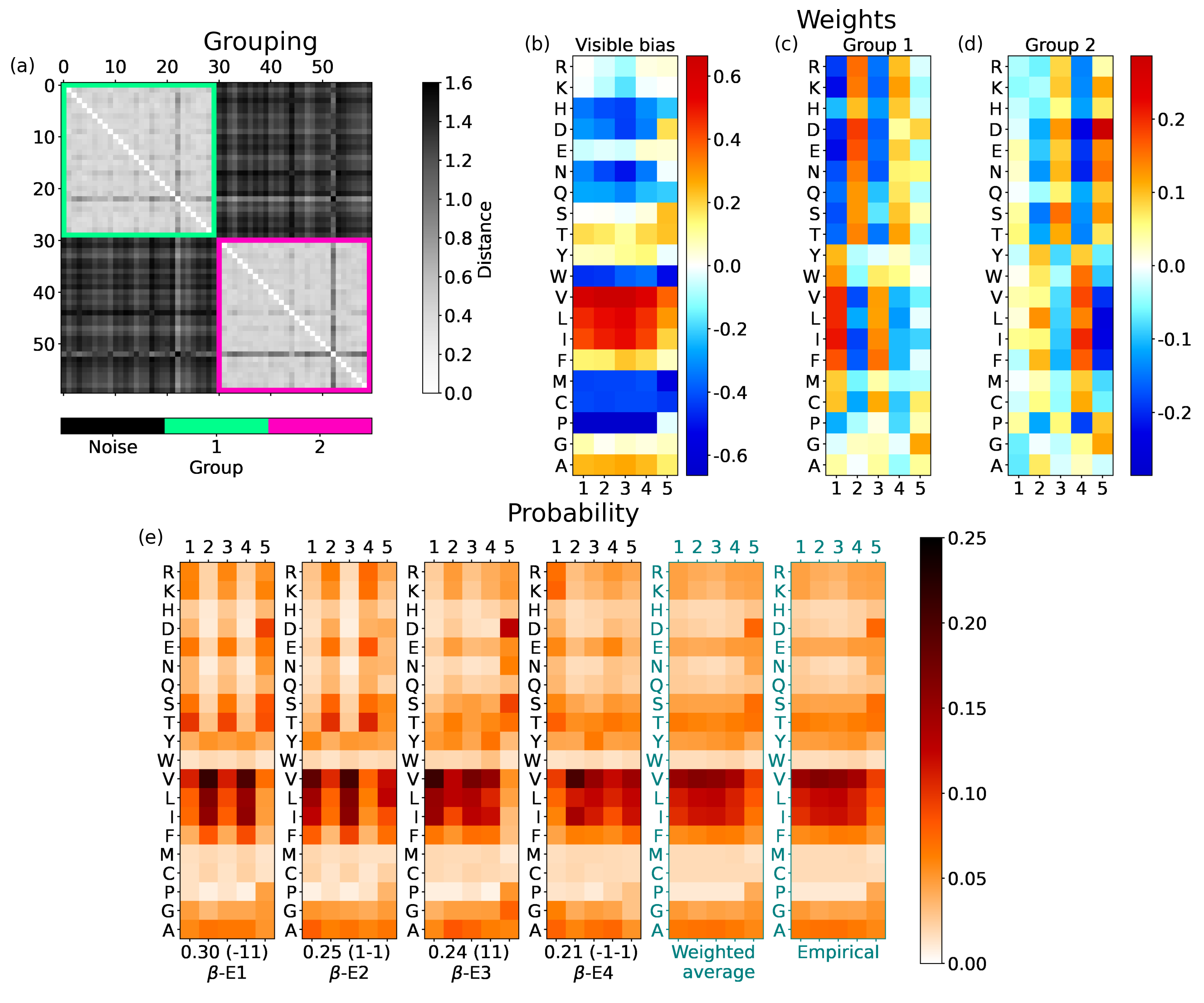}
 \caption{For the end of $\beta$-sheets with 2 hidden units, the same scheme as in Fig.~\ref{fig:a1}.  Hidden bias: $b_1=-0.349$, $b_2=-0.226$.}
 \label{fig:b2}
\end{figure*} 

The weighted average of probabilities for $\beta$-S$1\ldots 4$, as before for $\alpha$-helices, matches the empirical distributions (last panels of Figure~\ref{fig:b1}(e)).
The RBM-learned decomposition thus splits the start of $\beta$-sheets into four modes: the first two modes promote amphiphilic patterns, the last two modes favor uniform stretches of four $\mathbb N$'s (mostly I, L, V) capped by a different type of amino acid. 
This decomposition somewhat joins previous results in which the amphiphilic alternation of $\beta$-sheets was seen by different works, with more straightforward statistical tools, either over-represented~\cite{mand02} or under-represented ~\cite{broo00}.

The bias $a_i$ at the start of $\beta$-sheets shows (see Figure~\ref{fig:b1}(b)) a uniform distribution of amino acids at different positions of the chain. For instance, aliphatic amino acids show a high bias. However, for a small subset of amino acids, there emerges variability. For example, Arg (R) and Lys (K) have a decreasing bias from the first position in the $\beta$ strand to the following ones. Perhaps the most interesting behavior is observed for Gly, with a high bias except at the second position $\gamma=2$, suggesting that Gly is not likely to appear there.

\subsection{End of $\beta$-sheets}
\label{sec:b2}

Generally, the analysis of the end of $\beta$-sheets retraces the start of $\beta$-sheets. Thus, on average, the ensemble of RBMs can capture only patterns of little complexity in $\beta$-sheets compared to those of $\alpha$-helices.
We take $N_h=2$ also for the end of $\beta$-sheets, and again we observe two groups similar to those at the start of $\beta$-sheets (Figure~\ref{fig:b2}(c), (d)). 

Visible biases (Figure~\ref{fig:b2}(b)) show a uniform distribution of amino acids at different positions close to the end of $\beta$-sheets. However, there is a significant increase of the bias at the last position $\gamma=5$ for many small amino acids. Furthermore, many of these are polar (Asp (D), Asn (N), Ser (S), Thr (T)), and there is also Pro (P). In the next section, we will stress that Pro is often positively correlated with polar amino acids.
The biases shown in Figure~\ref{fig:b2}(b) are different from those we find at the start of $\beta$-sheets (Figure~\ref{fig:b1}(b)). As a consequence, the probabilities in Figure~\ref{fig:b2}(e) diverge slightly from those in Figure~\ref{fig:b1}(e). In particular, mode $\beta$-E3 promotes sequences as $\mathbb{NNNNP}$, $\mathbb{NNNN}$(Pro), or $\mathbb{NNNN}$(Gly) 

\subsection{Amino acid similarities}

The abundance or absence of a given amino acid in $\alpha$-helices or $\beta$-sheets is primarily encoded in the visible biases $a_i$. One can check that they correlate with results from standard statistical analysis~\cite{malk08}. However, these biases are not directly related to the polarity or size of amino acids. Hence, they do not provide complete information on the amino acid patterns in secondary structures.

The refined information on amino acid similarities is given by the weights shown in panels (c), (d), and eventually (e) of Figures~\ref{fig:a1},~\ref{fig:a2},~\ref{fig:b1}, and~\ref{fig:b2}. 
Each row in a panel shows the weights of a given amino acid in that group.
The similarity of amino acids in a group emerges when their weights are interchangeable, i.e., the $\Gamma=5$ weights appearing in the row of a given amino acid can be swapped with the other ones in a row of an equivalent amino acid without a significant change in the whole set of weight $w_{ij}$ of the corresponding hidden unit $j$.

\begin{figure*}[t!]
 \centering
 \includegraphics[width=0.9\textwidth]{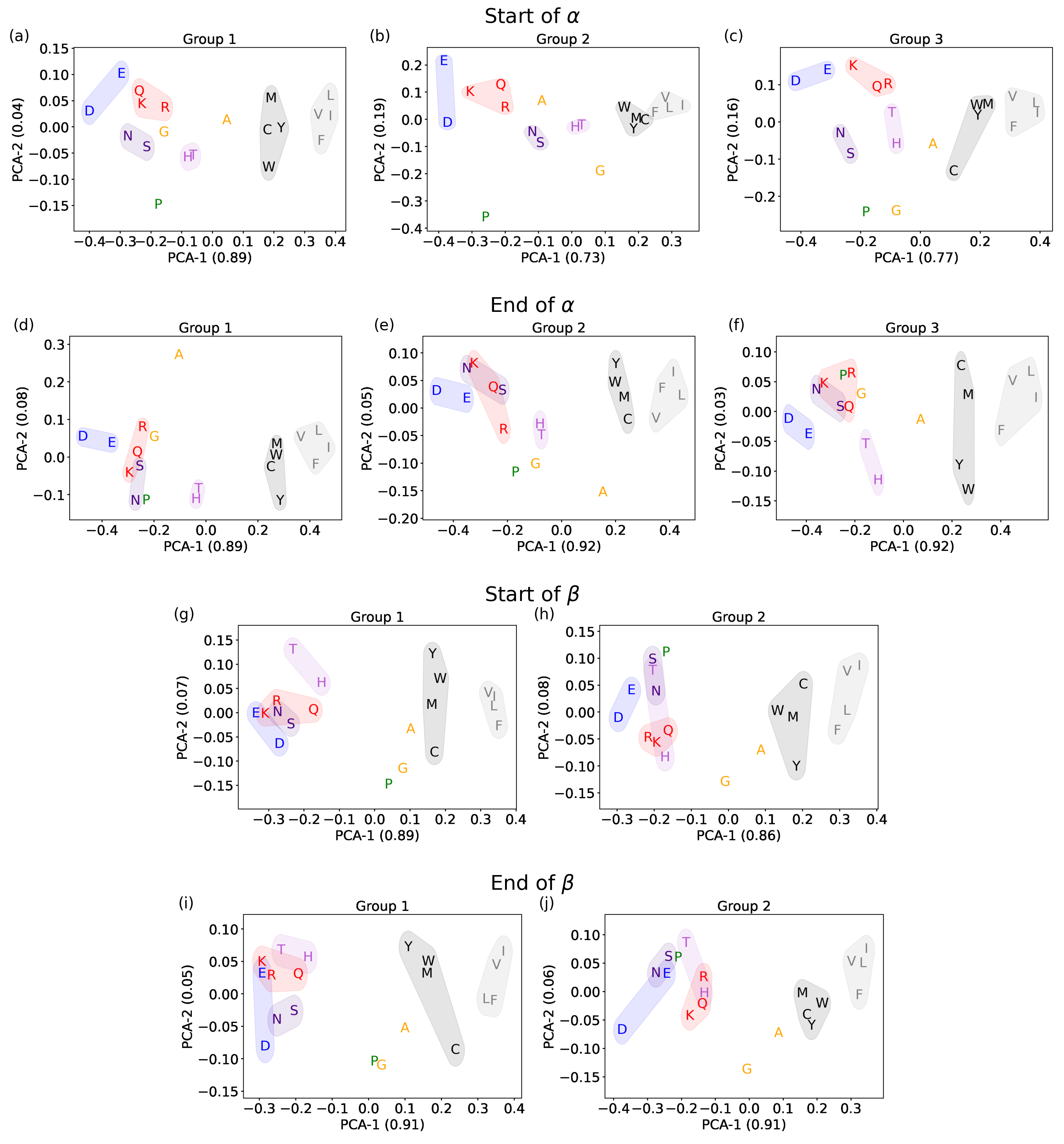}
 \caption{Principal component analysis of amino acid weights: each panel shows the first two components of the PCA for each amino acid in a hidden-unit group for a given part of the secondary structure. Color-shaded ensembles and single amino acids are discussed in the text. 
 (a), (b), (c) are the groups at the start of $\alpha$-helices, shown in Figure~\ref{fig:a1}.
 (d), (e), (f) are the groups at the end of $\alpha$-helices (Figure~\ref{fig:a2}).
 (g), (h) are the groups at the start of $\beta$-sheets (Figure~\ref{fig:b1}).
 (i), (j) are the groups at the end of $\beta$-sheets (Figure~\ref{fig:b2}).
 }
 \label{fig:PCA}
\end{figure*} 

In our unsupervised machine learning approach,
the salient traits of amino acids' similarity emerge from the first two components of the principal component analysis  (PCA) applied to their weights.
For all groups shown in Figures~\ref{fig:a1},~\ref{fig:a2},~\ref{fig:b1}, and~\ref{fig:b2}, we show these two PCA components in 
Figure~\ref{fig:PCA}.
The number on the axes label represents the average variance {\em explained} by each PCA component, measuring its relevance. In all cases, the first component of PCA, PCA-1, explains the major part of the variance and is related to the polarity of amino acids.

Families of interchangeable amino acids emerge, as highlighted in all panels of Figure~\ref{fig:PCA}. Let us discuss our interpretation of these plots by collecting similar amino acids into small coherent groups. We define this by looking primarily at their PCA components in $\alpha$-helices, where there is a clearer subdivision. Our amino acid cataloging was anticipated in Table~\ref{tab:1}.

\paragraph{Aspartic acid (D), glutamic acid (E)} 
These negatively charged amino acids are always at the left boundary of the PCA-1 component. Looking at the general arrangement of amino acids in the panels of Figure~\ref{fig:PCA}, we interpret this as a signal of a strong hydrophilic tendency that stands out even among charged and polar amino acids. Indeed, in all cases, the Pearson coefficient between PCA-1 and hydrophobicity is $\approx 0.9$.
Although Asp and Glu seem mostly similar, for the end of $\beta$-sheets, in Figure~\ref{fig:PCA}(j), we observe that the Asp (D) stands away from other polar amino acids. This indicates that Asp has a special role at the end of $\beta$-sheets, which cannot be implemented even by Glu.

\paragraph{Asparagine (N), serine (S)}
The PCA always shows these two small polar amino acids very close to each other. Moreover, they are placed between the pair of Asp and Glu and the central part of the PCA-1 component. This should be related to their lower hydrophilic tendency.

\paragraph{Lysine (K), arginine (R), glutamine (Q)} 
These amino acids have positively charged (K, R) or polar (Q) long side-chains. They appear similar, and for them, we can retrace the comments just made for Asn and Ser.

\paragraph{Histidine (H), threonine (T)}
Histidine is a weakly positively charged, large ($>150$ Da) amino acid, while Thr is a polar, small ($<120$ Da) amino acid. Thus, it is  surprising to find them very well paired in the PCA plots for $\alpha$-helices, where they sit in a middle region and are not very close to other hydrophilic amino acids. 
Hence, His and Thr display a similar weak tendency to contribute to the amphiphilic pattern in $\alpha$-helices.
In $\beta$-sheets, instead, they are not so correlated and are more overlapped with other polar amino acids.

\paragraph{Tyrosine (Y), tryptophan (W), methionine (M), cysteine (C)}
These amino acids always have very similar PCA values on the right-hand side of the panels. This quartet comprises a duo of aromatic amino acids (Y, W) and a duo of nonpolar ones with sulfur (M, C). 
In particular, Cys is a small, unique amino acid that can form disulfide bonds. Yet, the PCA correctly places it in the mild hydrophobic region (i.e., with positive but not extreme PCA-1 values).

\paragraph{Valine (V), leucine (L), isoleucine (I), phenylalanine (F)}
Three aliphatic amino acids (V, L, I) and Phe are always equivalently set on the rightmost side of the PCA-1 component. Our analysis with RBMs thus reveals that these four amino acids should be regarded as the strongest hydrophobic amino acids.

\paragraph{Alanine (A)}
Ala shows neither a clear hydrophobic nor hydrophilic tendency in the PCA plots of $\alpha$-helices (Figure~\ref{fig:PCA}(a)-(f)). Nevertheless, we find a peculiar isolation of Ala from the other amino acids in group 1 and group 2 of the end of $\alpha$-helices (Figure~\ref{fig:PCA}(d) and (e)), with an opposite sign of the PCA-2 component in the two cases. As discussed above, this is related to the unique role of Ala in helices, in particular at their end, where stretches of five Ala are not rare. However, in $\beta$-sheets, Ala shows a mild tendency to cluster with the $\mathbb N$ group and thus behave as hydrophobic (Figure~\ref{fig:PCA}(g)-(j)).

\paragraph{Glycine (G)}
Even if Gly is a nonpolar amino acid, in $\alpha$-helices, it is mainly found in the region populated by hydrophilic amino acids. However, this is not the case in $\beta$-sheets, where Gly is not affiliated with other groups.

\paragraph{Proline (P)}
Similarly to Gly, Pro is not polar but is often aligned with polar amino acids along the PCA-1. However, P displays several extreme values of the PCA-2, which isolate it from the other amino acids. The most striking case is in group 2 at the start of $\alpha$-helices (Figure~\ref{fig:PCA}(b)), which RBMs use to highlight the importance of Pro in this portion of the secondary structure.

Before concluding, we note that our PCA plots are similar to the embedding learned by much more complex neural networks using Transformers \cite{rives2021biological}. That analysis showed that the machine catalogs amino acids based on their biological properties.

\section{Conclusions}

We introduce and showcase how an ensemble analysis of (unsupervised) machine-learning models, based on restricted Boltzmann machines (RBMs) and with an information bottleneck in encoding data correlations, offers relatively easy reading of precise yet unexpected similarities between amino acids and emphasizes essential features for building secondary structures. Besides recovering a way to promote the frequent amphiphilic design of $\alpha$-helices and $\beta$-sheets, RBMs discover that there are relevant motifs that, to the best of our knowledge, were not known. 

The most diverse scenario is at the start of $\alpha$-helices. RBMs recover the known relative abundance of Pro in their first positions and promote it to the role of a highly relevant feature in addition to amphiphilicity. Moreover, RBMs add information on correlations between Pro and other amino acids, particularly Asp and Glu, which leads to two typical types of helices starting with Pro. Our complete analysis reveals a frequent alignment of Pro with polar amino acids.

At the end of $\alpha$-helices, there emerges a particular behavior of Ala, which is the distinguishing amino acid between two otherwise similar amphiphilic patterns. This bimodality implies that in nature there is a class of $\alpha$-helices closed by stretches richer in Ala than in typical helices.

Moreover, our analysis allows refining the separation between polar and nonpolar amino acids, highlighting intriguing subclasses. The most unexpected is the coupling of His and Thr in $\alpha$-helices, where they do not contribute to the amphiphilic patterns. Then, for instance, we find the coupling of Phe with the aliphatic amino acids or the alignment of Trp with Tyr, Met, and Cys. 

The first component of our PCA (PCA-1) is strongly correlated but does not follow precisely the ranking of hydrophobicity reported in the literature. Nevertheless, PCA-1 explains most of the fluctuations of weights in the RBM. Hence it is crucial to unveil its meaning. We conjecture that PCA-1, the main feature learned by RBMs to reproduce realistic alternations of polarity in secondary structures, expresses a form of {\em effective hydrophobicity}. In other words, it reveals how much an amino acid, in $\alpha$-helices and $\beta$-sheets, is mainly focused on the role of being either hydrophobic or hydrophilic. For example, Asp most often displays the strongest negative PCA-1 value (and has a special role in closing $\beta$-sheets), even if it is not the most hydrophilic amino acid.

To conclude, the RBM is a simple unsupervised machine learning method that retrieves known results and enriches previous knowledge. Moreover, the RBM's architecture is readable and, with some effort, interpretable, yielding nontrivial information inaccessible by standard statistical tools. For example, we have provided an interpretation of the RBM weights in our study of amino acid patterns and similarities in secondary structures. 
However, the richness of the results may allow the reader to notice additional details of the arrangement of amino acids in the protein's secondary structures. 

\paragraph*{Acknowledgments}

The authors thank Aurelien Decelle, Emanuele Locatelli, Lorenzo Rosset, and Antonio Trovato for the useful discussions and feedback on the text.
MB and EO are supported by research grants BAIE$\_$BIRD2021$\_$01 and ORLA$\_$BIRD2020$\_$01 of the University of Padova.


\renewcommand{\thefigure}{S\arabic{figure}}
\renewcommand\theequation{S\arabic{equation}}
\renewcommand{\thesection}{S\arabic{section}}
\setcounter{figure}{0}
\setcounter{equation}{0}

\clearpage
\section*{Supplementary Information}

We provide additional details on the restricted Boltzmann machines (RBMs) and on the clustering procedure.

Our RBMs are trained with well-known optimizations~\cite{hinton2012practical,fischer2014training} and with the "centering trick"~\cite{tang11,mont12}.
We build the ensemble of RBMs, with a fixed number of hidden units $N_h$, by training $R$ different realizations. Each realization is characterized by the RBM's random state, which determines the weights initialization and the data split into training (80\%) and validation (20\%) sets. Thus, we obtain a set of RBMs differing for parameter values and slightly for analyzed datasets. 

The bipartite structure of the RBM allows storing the information within weights and biases in many ways due to the invariance for permutation and sign reversal of hidden units. To overcome this variability when comparing units, we isolate each hidden unit $j$ in an RBM and compare its weights $w_{ij}$, and those of its mirror image $-w_{ij}$, with those of all other hidden units in the same RBM and other RBMs of the ensemble. Thus, for every pair of hidden units $j,m$, we compute the minimum Euclidean distance among them or their mirror versions, 
\begin{equation}
 d_{jm}=
 \min
 \left[
 \sqrt{\sum_i |w_{ij} - w_{im}|^2},
 \sqrt{\sum_i |w_{ij} + w_{im}|^2}
 \right].
\end{equation}
We then feed the distance matrix $d_{jm}$ to a popular density-based algorithm, DBSCAN~\cite{birant2007st,mehta2019high}, to perform clustering. We focus on tuning two parameters: a radius around each data point ($\epsilon$) and the minimum number of samples ($\min_s$) within $\epsilon$ from a data point that would prevent its labeling as {\em noise}, i.e., that would grant to put that point in a cluster.

\begin{figure}[b!]
 \centering
 \includegraphics[width=0.43\textwidth]{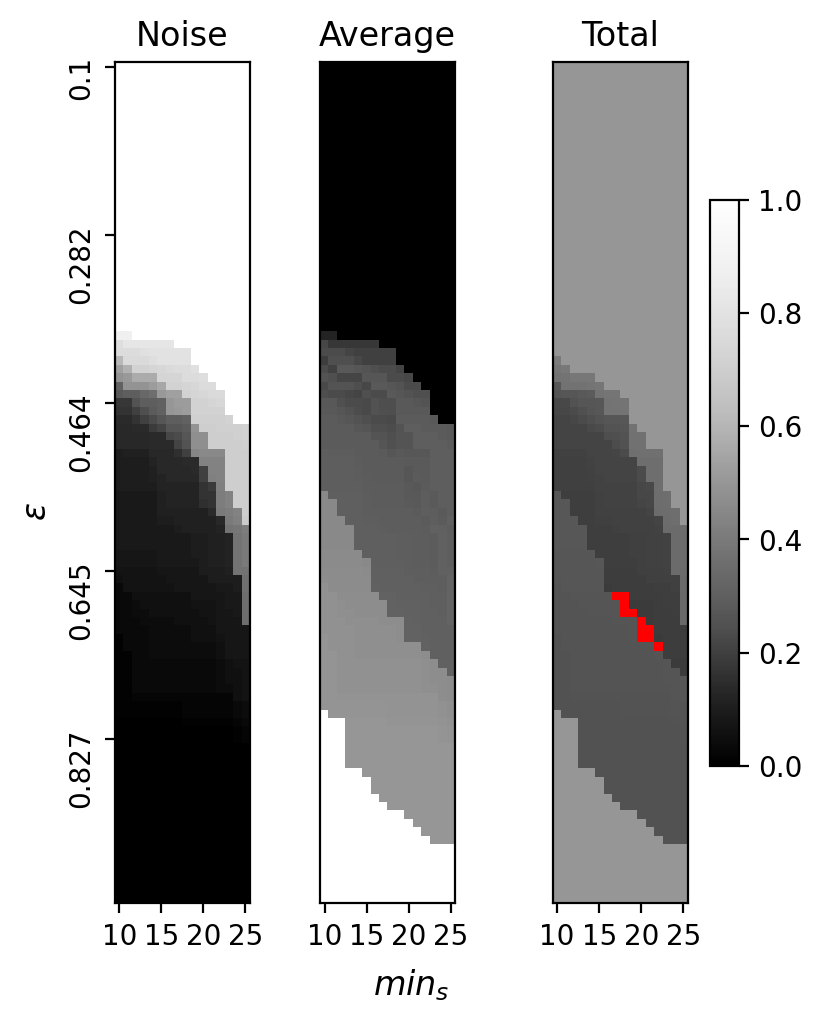}
 \caption{Example of parameter tuning in DBSCAN for the start of $\alpha$-helices. The "Noise" matrix represents the fraction of noise points (first term in \eqref{dbscantune}). The "Average" matrix represents the average group size divided by the total number of hidden units in the ensemble (second term in \eqref{dbscantune}). The "Total" matrix is the sum of the previous two and coincides with the cost function in \eqref{dbscantune}. The red points highlight the optimal region of the parameters.}
 \label{fig:dbtune}
\end{figure}

For tuning $\epsilon$ and $min_s$, we introduce a cost function $C(\epsilon,min_s)$ whose minimum corresponds to the optimal parameter values,
\begin{equation}
 C(\epsilon,min_s) = \left[1- \sum_{g \in \mathcal{G}} \frac{\Omega (g)}{R\, N_h}\right]+
 \left[ \frac{\langle \Omega (g) \rangle_{g \in \mathcal{G}}}{R\,N_h}\right]
 \label{dbscantune}
\end{equation}
where $\mathcal{G} := \mathcal{G}(\epsilon,min_s)$ denotes the set of groups returned by DBSCAN and $\Omega (g)$ denotes the number of hidden units in group $g \in \mathcal{G}$. 
The first term in \eqref{dbscantune} is the fraction of hidden units that are cataloged as noise, hence it penalizes configurations with high noise. The second term is proportional to the average cluster size $\langle \Omega (g) \rangle_{g \in \mathcal{G}}$ and penalizes configurations with all the hidden units merged in a single, giant cluster. In Figure~\ref{fig:dbtune} we show the results and the intermediate steps of the parameter tuning at the start of $\alpha$-helices. For the other cases, the results are similar.
The procedure returns an optimal region of parameters: we choose the average points within this region as optimal parameters.

The average RBM is then built as the average of weights $w_{ij}$ and biases $b_j$ within each group after aligning all its hidden units to minimize the distance from a reference one. 
For better overall visualization, since aliphatic amino acids (I, L, V) always yield a well-defined pattern, we adopt the convention that their weights at position $\gamma = 1$ are positive.
The visible bias $a_i$ is instead independent of the grouping; hence, it is averaged among all the original RBMs.

For each hidden unit $j$ representing the average of units in a group, we analyze weights $w_{ij}$ to extract the similarity among amino acids. For this purpose, we split the array of weights $w_{ij}$ into $20$ sub-vectors of length $\Gamma$ (i.e., a row if $w_{ij}$ is reshaped to a $20\times \Gamma$ matrix), each one related to a specific amino acid. To extract the most relevant linear combinations of coordinates in the $\Gamma$-dimensional space of sub-vectors in a group, we perform a principal component analysis (PCA)~\cite{mehta2019high} on them.
The PCA ranks the most relevant and independent linear combinations of coordinates in the $\Gamma$-dimensional space.


%

\end{document}